# Measurement of the Spin Relaxation Time of Single Electrons in a Silicon MOS-Based Quantum Dot

M. Xiao, M. G. House, and H. W. Jiang

Department of Physics and Astronomy, University of California at Los Angeles
Los Angeles, CA 90095

We report on measurements of the spin relaxation time $T_1$ of individual electron spins in the few electron regime of a $Si/SiO_2$-based quantum dot (QD). Energy-spectroscopy of the QD has been performed using a charge sensing technique. The spin relaxation times are subsequently measured in the time-domain by a pump-and-probe method. For the QD that contains an unpaired spin, likely only a single electron, we find that $T_1$ depends strongly on the applied magnetic field. Possible mechanisms leading to the observed spin relaxation are discussed.

The potential of using individual electron spin states in a semiconductor quantum dot for quantum information processing has triggered a stream of experimental investigations in recent years to detect and manipulate single spins in the few electron limit.[1] The spin relaxation time $T_1$, an important measure of the interaction between a two-level quantum system and its environment, has now been successfully measured in GaAs based quantum dots for spin-flip transitions between two magnetic field induced Zeeman sublevels,[2,3] and between a singlet ground-state and a triplet excited state,[4,5,6] using electrical pump-and-probe methods. Similar manipulation/measurement in Si based quantum dots has been long awaited, as isotopically pure Si materials are predicted to have extremely long phase-coherence times. However, comparable electrostatically-defined Si devices, which have only started to emerge very recently, usually do not have the stability and/or controllability required for the demanding spin relaxation measurements. Here we present such a $T_1$ measurement of individual electron spins in the few electron regime of a Si MOS-based quantum dot.

To conduct the experiment, we fabricated a collection of Si MOS-based quantum dots (see methods) after almost 3 years of developmental effort. Quantum dots (QD) fabricated in this manner show excellent stability and controllability, which allow us to study the tunneling dynamics of individual electrons and to measure the spin relaxation time $T_1$ in the few electron regime. The pattern of the confinement gates is shown in the electron micrograph in Fig. 1a. The cross-sectional view of the device is shown in Fig. 1b. This double insulating layered Si MOS based quantum dot structure is architecturally similar to those reported in a number of earlier publications[7] and should be fully compatible with commercial Si MOS technology.

To form the QD and the read-out channel, the top global gate was set at a fixed positive value such that the 2D electrons ($\sim 3 \times 10^{11}/cm^2$) could accumulate near the interface of the Si and $SiO_2$ across the entire sample. Lower voltages applied to the five gates LT, RT, LB, RB and P shaped the QD. While gate P was used primarily to control the number of electrons in the QD, gates LT-LB and RT-RB controlled the transparency of the left and right barriers, respectively. Gate Q, along with RT and LT, formed a 1D charge sensing channel to count the number of electrons via capacitive coupling. Since the global top gate is only 150 nm above the QD, and the side gates only 50 nm, the capacitive coupling between the QD and the charge sensing channel was relatively weak as electric field lines could be screened out by the metallic gates. A small gap between TL and TR was thus created to maximize this capacitive coupling while there was no direct electron tunneling between the QD and the sensing channel.

We can detect the addition of an electron to the QD by tracking changes in the 1D channel current. Typically the addition of a single electron would result in a reduction in the total current of about 1%. To offset the large current background, we used a lock-in detection method developed earlier for GaAs work[8]. A square shaped pulse was superimposed on the DC bias on P. A lock-in detector in sync with the pulse frequency measured the change of the channel current due to the pulse modulation. Fig. 1c shows a typical trace of the lock-in signal as a function of the voltage applied to gate P. The four dips indicate the transitions in the charge states by addition/subtraction of single electrons. The QD was put into an environment in which the left barrier was essentially opaque and the tunneling frequency between the right barrier and the reservoir was about 1 KHz. The tunneling rate can be tuned continuously from 100Hz to 30 KHz for the last few electrons and can be measured in time-domain by an oscilloscope. The four dips shown are most likely the last 4 electrons in the QD, as we couldn't detect any additional dips as the QD was further squeezed. We verified that the absence of the additional peak was not due to the closure of the QD by increasing the voltages on RT and RB (i.e., the transparency of the right barrier). In any case, for convenience, we use N=1, 2, 3, 4 throughout the paper for assignment.

The information contained in the signal goes beyond simple charge counting. For instance, varying the pulse amplitude can reveal excited states. Fig. 2a and 2b show in a gray scale plot the derivative of the signal as a function of the pulse amplitude and gate voltage for the N=0 ↔ N=1 and the N=1 ↔ N=2 transitions, respectively. Ignoring the interior bright line for a moment, a triangular pattern can be visualized. The left line is due to the front edge of the pulse beginning the process of electron loading while the right line is for the point where the ground state electron is unloading. In addition, for both plots, there is an extra interior line that indicates an excited state. The excited state becomes visible when the excitation frequency is high enough in comparison to the relaxation rate from the excited state to the ground state. As shown in the energy diagrams, a pulse with sufficiently high amplitude can populate either the ground state or the excited state during the high-voltage cycle and depopulate during the low-voltage cycle. We found that the interior line terminates on the right side for the 0-1 transition and terminates on the left side for the 1-2 transition.

We studied the dependence of the termination points on a magnetic field applied parallel to the Si/SiO$_2$ interface and found that point A was largely independent of the magnetic field while the termination point B varied linearly with the field, as shown in Fig. 2c. Following arguments from excited state spectroscopy[1], point B measures the energy difference between one of the excited states of N=1 and the ground state of N=1. Since it depends linearly on field, it is most likely a measure of the spacing of the two Zeeman sublevels. Likewise, point A also measures the energy difference between one of the excited states of N=1 and the ground state of N=1. Since the spacing is independent of magnetic field, we assign it to the orbital level spacing of the QD. Assuming the g-factor of an electron in Si to be 2, a conversion factor between the pulse voltage and the energy can be extracted from the magnetic field dependence of the energy level spacing to be 27 meV/V. This conversion factor is consistent with that obtained from the transport measurement of the Coulomb diamonds. The *B*-independent energy spacing is therefore about 0.4 meV. The Coulomb charging energies needed to add an additional electron to N=1, N=2, and N=3 QD are 5 meV, 3.8 meV, and 3 meV, respectively.

Having established the energies of the QD charge states, we now present a relaxation measurement for both the N=1 and N=2 QD. For the N=1 QD, it should be a spin-flip transition between two magnetic field induced Zeeman sublevels, as discussed. We used a two-step pulse sequence, adapting again from earlier GaAs work[2,3,5]. Fig. 3 illustrates schematically the working principle of this electrical pump-and-probe technique. The QD during the first phase is emptied or initialized as the Fermi level sits below the ground state. The Fermi level is then moved above both the down-spin ground and the up-spin excited states during the second phase. After waiting for a certain period of time $t_W$, the Fermi level is placed to the middle of the two levels for the state read-out during third phase. For $t_W \gg T_1$, the electron is expected to be relaxed to the ground state and there should be no tunneling from the QD to the reservoir. However, for short $t_W$, if the electron still remains in the excited state, the electron tunnels out and subsequently tunnels back into the ground state, generating a transient signal in the charge-sensing channel. In principle, this tunneling event can be detected by applying a single pulse (i.e., single-shot measurement)[2]. However, the relatively poor signal-to-noise ratio of our detection, about 1:5, prevented us from seeing the tunneling event in real-time. We therefore applied multiple pulses and averaged the channel signal over several thousand times. The resulting signal was captured by a digital oscilloscope. The sample oscilloscope trace displayed in Fig. 3 (d) shows a broadened peak. The broadening is expected from the statistical distributions of the tunneling electron in and out of the QD as simulated in Fig. 3 (c) (see the methods section).

In Fig. 4(a) the tunneling peak is shown for several waiting times at B=4T. The trend of the reduction of the height with increasing waiting time can be clearly seen. This dependence is plotted in Fig. 4 (b) and was fit to an exponential decay $A = \exp\left(-\frac{t_W}{T_1}\right)$ to extract $T_1$. Our measurement capability is limited by the charge detector bandwidth at the short-time scales. The relaxation rates from the excited state to ground state are plotted as a function of the magnetic field. For the N=1 QD the relaxation rate shows a strong dependence on B. In contrast, the relaxation time is essentially a constant of around 5 mSec for the N=2 QD.

For a N=1 QD, in the presence of a magnetic field, the main mechanism for electrons to relax from one Zeeman split sublevel to another is through the stochastic electric-field fluctuations caused by phonons of the host materials[1,9,10,11,12,13]. The coupling between the magnetic fields of spins to the electrical fields of phonons can be facilitated by the relatively strong spin-orbital coupling (SOC) in semiconductors. For GaAs QDs, it has been demonstrated that the spin relaxation rate $T_1^{-1}$ depends both on the piezoelectric effect (i.e., piezoelectric phonons) and the spatial deformation of the crystal structure (i.e., deformation potential phonons).[1,3,4,5] Since it is not a polar crystal, for Si the piezoelectric phonon contribution should be negligibly small[11], so the relaxation is expected to be dominated by the deformation potential of acoustic phonons. The relaxation rate depends on the phonon density of states at the Zeeman energy, the amplitude of the electric fields generated by the phonons, and the strength of the SOC. The fingerprint of the deformation potential is that $T_1^{-1}$ is expected to depend on the seventh power of the magnetic field in the limit that the wavelength of the phonon is larger than the size of the QD.[1,11]

If we fit data preferentially to $B^7$, a proportionality constant of 0.017 ($T^7$-sec) can be extracted. Since there is no theory that works out specifically for electron spins near the Si/SiO$_2$ interface, where the SOC is known to be dominated by the Rashba effect due to the strong electric field, we cannot further discuss this proportionality constant quantitatively. The data also clearly shows a tendency for weaker B dependence as B→0. This residual relaxation could be caused by electrical noise generated by background charge fluctuations, as the noise can dominate when the contributions from phonons rapidly diminish as B→0. On the other hand, we cannot rule out the possibility of a different spin relaxation mechanism, which becomes more prominent as B→0. We also do not know what influence the presence of the Si/SiO$_2$ interface has on spin relaxation. The above discussion has focused on theory that was developed for bulk phonons, but spin relaxation might be impacted by phonon scattering from the interface. Further theoretical study of the spin relaxation mechanisms in this type of QD is well deserved. It should also be noted that during the preparation of this manuscript, we have received a preprint from HRL Laboratories, LLC that reported a measurement of spin relaxation time of Si/SiGe quantum dots.[14] We cannot however make a quantitative comparison with the data since the two material systems are different.

In contrast to the N=1 case, the relaxation time measured is roughly independent of the magnetic field around 5 mSec for N=2. Since the energy splitting between two valleys in Si is generally unknown for QD structures,[15] we do not know the exact excited states spectra to positively identify the origin of the relaxation transition, even for a two-electron QD. Normally, for a QD with paired spins, the spin relaxation involves a transition between one of the triplet-states and a singlet-state.[1] The relaxation rate in this case should be proportional to the phonon DOS at the transition energy. If we use the B-independent level spacing of 0.4 meV to roughly estimate the spin flipping energy, it is equivalent to a Zeeman energy at 3T not too far off from the observed cross-over of B=4T.

Here we would like to re-emphasize that while we conveniently label the electron numbers by N=1, 2 etc., we cannot be 100% certain that the QD indeed contained only the last two electrons. However, in any case, the discussion above should apply to both the cases of an unpaired electron and paired electrons respectively.

In conclusion, we have measured the spin relaxation times of a few-electron Si/SiO$_2$-based QD and studied their magnetic field dependence by an electrical pump-and-probe. Given the prominent importance of Si/SiO$_2$ based materials in mainstream electronics, it is crucial to determine whether the Si/SiO$_2$ QD is indeed a good candidate for quantum information processing. An interesting future experiment will be to measure the phase coherence time T$_2$ to see if the promise of long phase coherence time for bulk Si can be kept in the presence of a Si/SiO$_2$ interface.

The authors would like to thank Andy Hunter and Mark Gyure for bringing our attention to the multiple pulse averaging technique and Andrey Kiselev for useful discussions on spin relaxation mechanisms. The work is sponsored by U.S. Department of Defense. The views and conclusions contained in this document are those of the authors and should not be interpreted as representing the official policies, either expressly or implied, of the U.S. Government.

## Methods

The Si quantum dot device fabrication started with a commercial Si/SiO$_2$ wafer with a 50nm thick thermal oxide. First, multiple confinement gates were fabricated by electron-beam-lithography on the wafer, electrostatically defining a quantum dot. Then, a thin 100nm isolation Al$_2$O$_3$ layer was grown epitaxially by atomic layer deposition. Finally, a global gate was fabricated on top of the isolation layer for controlling the number of electrons to be accumulated in the quantum dot. Additionally, Ohmic-contacts, which connected to the 2D electron reservoir, were formed by implanting phosphorous ions at 45 keV at a dosage of 2x10$^{15}$/cm$^2$.

To sense the charge in the QD, a DC biasing voltage of 0.8 mV was applied to one contact of the channel while the resulting current was collected from another contact and amplified by a high-bandwidth (200 KHz) and low-noise (130 fA/√Hz) current amplifier (FEMTO DLPCA-200) at a gain of 10$^8$ V/A. The resistance of the sensing channel was set by gate Q to about 10$^5$ ohms, which corresponds to a bandwidth of 50 KHz, for the detection. Also, a Stanford Research System SRS535 Pulse/Delay Digital Generator was used to provide the electrical pulses and a SRS 830 lock-in amplifier was used to record time integrated signal. An Agilent Infiniium 54855A oscilloscope with a sampling rate up to 5Gs/Sec was used for the time-resolved measurements. The experiments were done in an Oxford top-loading He3 refrigerator with a base temperature of ~300 mK. A magnetic field was applied along the plane of the device. An in-plane field is used to introduce the Zeeman sublevel splitting, while minimizing the B-dependence of the orbital states.

The simulation of the read-out signal (i.e., the tunneling peak) in Fig. 3(c) was done using a rate equation model. The signal, when averaged over many periods of the pulse sequence, is proportional to the expectation value of the number of electrons on the quantum dot. We have developed a rate equation model to determine this probability as a function of time during the pulse sequence. The quantum dot is modeled as having three states: electron spin up, electron spin down, and no electron. The probabilities that the system is in each of the three states are contained in the vector $\mathbf{p} = (p_\uparrow, p_\downarrow, p_0)$, which evolves in time according to the equation

$$\frac{d}{dt}\mathbf{p} = Q\mathbf{p}$$

where $Q$ is a transition matrix that describes the instantaneous transition rates between the states. The transition rates depend on the dot potential, so there are three different transition matrices, one for each phase of the three-step cycle:

$$Q_I = \begin{pmatrix} -W - \Gamma_{\uparrow,out} & 0 & 0 \\ W & -\Gamma_{\downarrow,out} & 0 \\ \Gamma_{\uparrow,out} & \Gamma_{\downarrow,out} & 0 \end{pmatrix}$$

$$Q_{II} = \begin{pmatrix} -W & 0 & \Gamma_{\uparrow,in} \\ W & 0 & \Gamma_{\downarrow,in} \\ 0 & 0 & -\Gamma_{\downarrow,in} - \Gamma_{\uparrow,in} \end{pmatrix}$$

$$Q_{III} = \begin{pmatrix} -W - \Gamma_{\uparrow,out} & 0 & 0 \\ W & 0 & \Gamma_{\downarrow,in} \\ \Gamma_{\uparrow,out} & 0 & -\Gamma_{\downarrow,in} \end{pmatrix}$$

where $\Gamma_{\uparrow,in(out)}$ is the tunneling rate into (out of) the $|\uparrow\rangle$ state, $\Gamma_{\downarrow,in(out)}$ is the tunneling rate into (out of) the $|\downarrow\rangle$ state, and $W$ is the rate of relaxation from $|\uparrow\rangle$ to $|\downarrow\rangle$. These three matrices correspond to the initialization/reset, injection and wait, and read-out phases of the cycle, respectively. Each full pulse period the probability vector evolves according to

$$\mathbf{p}(t_I + t_{II} + t_{III}) = \exp(Q_{III}t_{III})\exp(Q_{II}t_{II})\exp(Q_I t_I)\mathbf{p}(0).$$

In the steady state $\mathbf{p}(t_I + t_{II} + t_{III}) = \mathbf{p}(0)$, so the steady state $\mathbf{p}(0)$ is an eigenvector of the matrix $\exp(Q_{III}t_{III})\exp(Q_{II}t_{II})\exp(Q_I t_I)$ with eigenvalue 1. In the limit $t_I \to \infty$, the electron always tunnels out during the empty phase, so the probabilities approach $\mathbf{p}(0) = (0,0,1)$. This is ideal since having no electron present at the beginning of the injection phase will maximize loading of the spin up state, but might not always hold since for practical reasons we must keep $t_I$ finite. The probability vector during the read-out phase is

$$\mathbf{p}(t) = \exp[Q_{III}(t - t_{II} - t_I)]\exp[Q_{II}t_{II}]\exp[Q_I t_I]\mathbf{p}(0), \qquad t_{II} \leq t \leq t_{II}+t_{III}.$$

The channel current *I(t)* is proportional to $p_\uparrow(t) + p_\downarrow(t)$, which during the read-out phase first increases, approximately like $\exp[-\Gamma_{\uparrow,out}(t - t_{II})]$, as spin-up electrons tunnel out of the quantum dot and then decreases, approximately like $\exp[-\Gamma_{\downarrow,in}(t - t_{II})]$, as spin-down electrons tunnel back in. It is this "tunneling peak" feature that signals the occupation of the spin-up state; its amplitude is reduced as we increase $t_{II}$ because spin-up electrons relax to spin-down during the injection phase. In the limit that the tunneling rates are much faster than *W*, the dependence on $t_{II}$ is $\exp(-Wt_{II})$, in which case we can determine the relaxation rate *W* by fitting the tunneling peak amplitude as a function of $t_{II}$ to an exponential decay curve. When *W* is on the same order of magnitude as the tunneling rates, as it sometimes is in our experiment, the rate equation model is useful for comparison to the observed data.

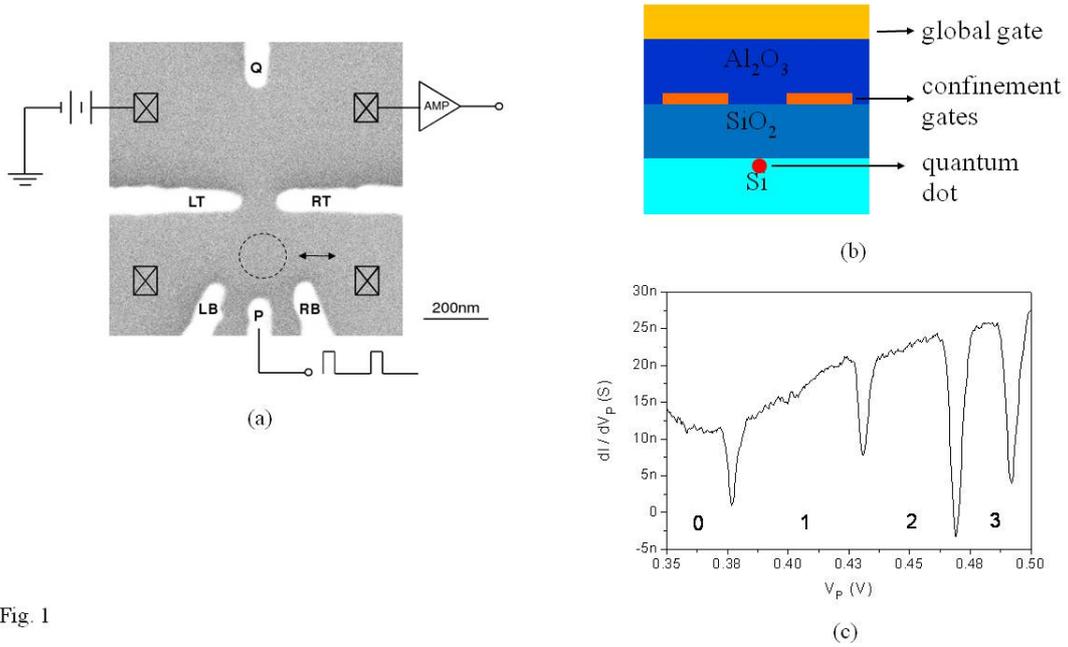

Fig. 1

Fig. 1 (a) Scanning electron micrograph of the confinement gates that define the QD (open circle), along with the measurement setup. (b) The cross-sectional view of the device. (c) A typical trace of the integrated current signal of the charge sensing channel as a function of the DC bias $V_P$, $V_{LT}$=-0.1V, $V_{RT}$=-0.16V, $V_{LB}$=-0.85V, $V_{RB}$=-0.44V, and $V_Q$==0.7V. A square pulse of $\delta V_p$=3 mV and f=10 KHz was superimposed to DC bias on gate P to dynamically charge and discharge the QD.

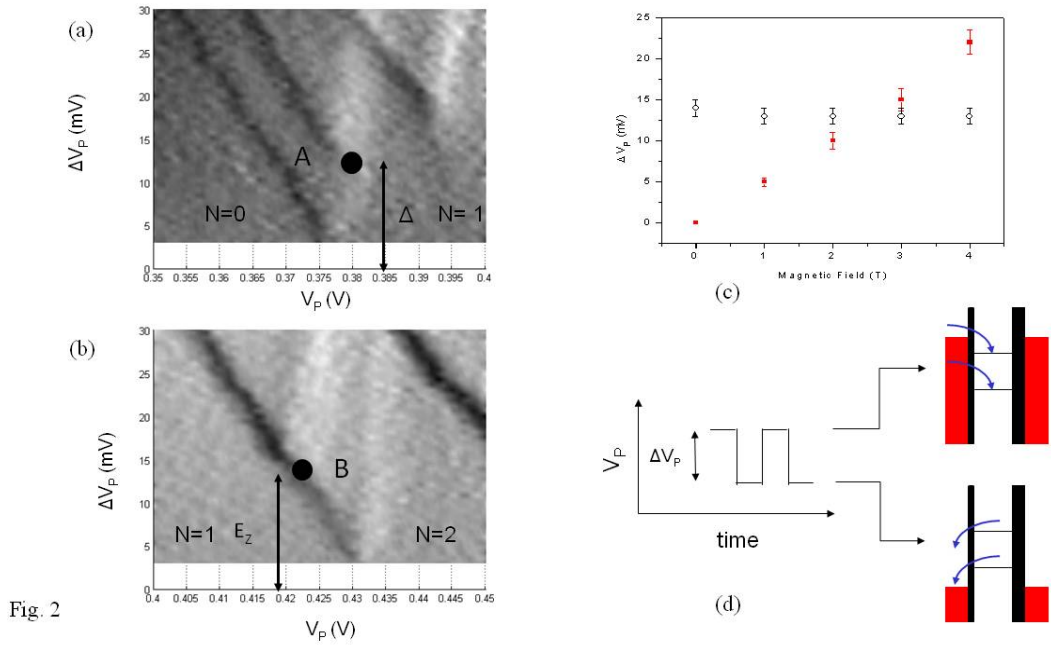

Fig. 2 Gray scale plots of the derivative of the current signal with respect to $\Delta V_p$ as a function of $\Delta V_p$ and $V_p$ at B=3T for (a) the N=0 to N=1 transition (open circles), and (b) the N=1 to N=2 transition (closed squares). (c) Pulse amplitude for the termination points of the excited state line, a measure of the QD level spacing and the Zeeman splitting, as a function of the magnetic field. (d) Schematic electrochemical potential diagrams illustrate the charging and discharging of the QD during the high and low voltages of the pulse.

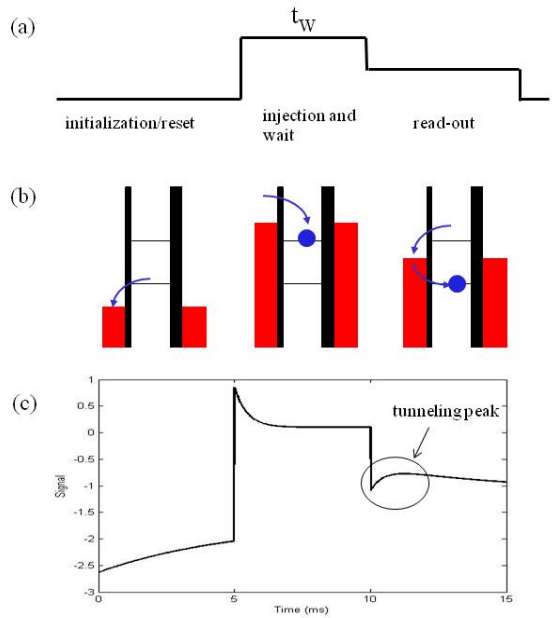

Fig. 3 (a) Two-step pulse sequence used for the T$_1$ measurement. (b) Schematic electrochemical potential diagrams for the three stages of the pulse. (c) Simulated current signal in time-domain, averaged over many cycles of such a pulse sequence. The simulation parameters used were $\Gamma_{\uparrow,in}=\Gamma_{\uparrow,out}=2$ kHz, $\Gamma_{\downarrow,in}=\Gamma_{\downarrow,out}=0.2$ kHz, $W=100$ Hz.

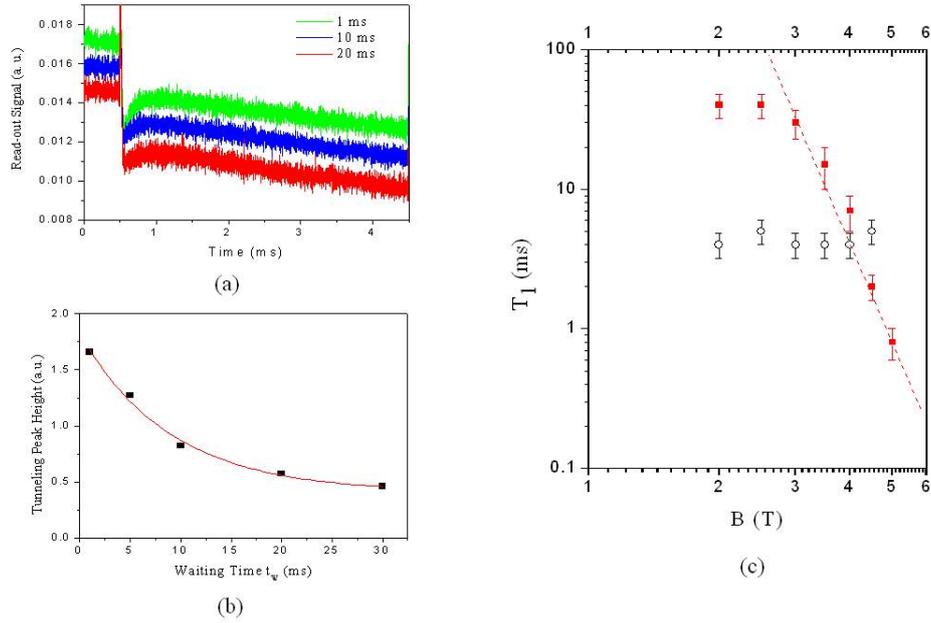

Fig. 4 (a) The tunneling signal at the read-out phase is displayed for three different waiting times. (b) The amplitude of the tunneling signal, proportional to the probability of occupying the spin-up state, as a function of $T_w$. The amplitude is fitted to an exponential curve that gives $T_1 \approx 10$ ms. (c) The relaxation times for the N=1 (closed squares) and N=2 (open dots) as a function of the magnetic field. The dashed line projects a $B^{-7}$ dependence.